\documentstyle[prl,aps,epsfig]{revtex}

\newlength{\textwidthm}
\begin{document}

\preprint{}
\title{How to Trap Photons? 
Storing 
Single-Photon Quantum States 
in Collective Atomic
Excitations \footnote{This paper is dedicated to Marlan O. Scully on the
occasion of his 60th birthday. We are grateful to him 
for introducing us to this exciting field and for 
his continuous inspiration and encouragement.
}}
\author{M.~Fleischhauer$^{1,2}$, S.~F.~Yelin$^3$, and M.~D.~Lukin$^1$}
\address{
	$^1$ ITAMP, Harvard-Smithsonian Center for Astrophysics,
      		Cambridge, MA~~02138 }
\address{$^2$ Sektion Physik, Ludwig-Maximilians-Universit\"at M\"unchen,
Theresienstr. 37, D-80333 M\"unchen, Germany}
\address{$^3$ Research Laboratory of Electronics, MIT, Cambridge, MA 02143}
\maketitle
\begin{abstract}
We show that it is possible to ``store'' quantum states of single-photon 
fields by mapping them onto {\it collective} meta-stable states of an
optically dense,
coherently driven medium inside an optical resonator. An adiabatic 
technique is suggested which allows 
to transfer non-classical correlations from 
traveling-wave single-photon wave-packets into 
atomic states and vise versa with nearly 100\% efficiency.
In contrast to previous approaches involving single atoms, 
the present technique does not require the strong coupling regime 
corresponding to high-$Q$ micro-cavities. Instead, 
intracavity Electromagnetically Induced Transparency is used to achieve 
a strong coupling between the cavity mode and the atoms.
\end{abstract}



\section{Introduction}


Nearly fifteen years ago Marlan Scully
and his co-workers envisioned that  
coherence effects in atoms can be used to correlate quantum 
fluctuations in lasers \cite{cel}. Since then the concepts of
atomic coherence and interference were extended and applied to many areas
of quantum optics and beyond \cite{scullybook}. 
Examples include electromagnetically induced transparency (EIT) \cite{eit},
lasing without inversion 
(LWI)~\cite{lwi}, quenching of spontaneous
emission~\cite{iwl}, sensitive 
spectroscopy in coherent media~\cite{magnetometry,lukin97prl}, 
and the enhancement of linear 
and nonlinear susceptibilities~\cite{index,nonlinear}.

The present contribution is stimulated by 
recent experiments, in which Electromagnetically Induced Transparency 
has been used to dramatically reduce the group velocity of 
light pulses in
a coherently driven, optically dense ensemble of atoms 
\cite{Hau99,Kash99,budker99}. 
This slow-down and the associated group delay can be viewed as 
a temporary storage of light energy in the 
atomic medium and its  subsequent release. The slowly 
traveling light pulses propagate, under ideal conditions,
without losses and distortion.  

The present paper demonstrates that it is possible to use 
closely related ideas to ``store'' and 
preserve {\it quantum states} of free-space light fields 
over a very long time interval. 
Processes of this kind open up  
interesting prospectives for quantum information processing without 
the usual ``strong coupling'' requirement of cavity QED.

An important class of schemes for 
quantum communication and computing in based on an elementary process in 
which  single quanta of excitation  are transfered back and forth 
between an atom and 
photon-number states of the radiation field 
\cite{Parkins93}. This is 
achieved within the framework of cavity QED by an adiabatic rotation 
of dark states \cite{STIRAP}
wherein a single atom is strongly coupled to 
the mode of a high-$Q$ micro-cavity. Based on this technique, 
excitations 
can be transferred from an atom in one cavity to a different atom in a 
second cavity, resulting in an entanglement of a pair of atoms 
separated by a long distance 
\cite{Cirac97,Enk97,Pellizzari97,Enk98}.
Also sources for
single-photon wave-packets referred to as photon guns \cite{Law97} or turnstile
devices \cite{Imamoglu94} were suggested and methods for entanglement 
engineering of single-photon wave-packets proposed \cite{Gheri98}. 
 Furthermore,  adiabatic passage 
of this kind can be used as  the basis for an elementary quantum logic 
gate \cite{Pellizzari95}. Experimental realizations of these ideas are however 
quite challenging, as the excitation rate determined by 
the vacuum Rabi-frequency (atom-cavity coupling constant) 
must exceed the decay out of the cavity. Despite an exciting 
progress  towards the realization of such a strong-coupling 
regime,  extreme technological challenges remain \cite{kimble}.  

The present proposal suggests an alternative root towards the 
solution of these 
problems. Specifically, we show here that it is possible 
to map the quantum states of {\it traveling light waves}
onto {\it collective} meta-stable states of 
optically dense, coherently driven media inside an optical resonator.
In particular, we suggest and analyze an adiabatic
transfer method which allows one to transfer non-classical 
states of light fields into 
atomic Zeeman sub-levels and vise versa with nearly 100\% efficiency. 
This process is based on the effect of
intracavity electromagnetically induced transparency, suggested  
in \cite{OL98}. 
In contrast to single-atom approaches, the technique described here, does 
not require the usual strong-coupling regime of cavity QED.  
The key mechanism
which allows us to avoid this stringent requirement is the use of an
optically dense many-atom system. In such a system single photons couple 
to {\it collective excitations} associated with EIT, and the corresponding 
coupling strength exceeds that of an individual atom by
the square root of the number of atoms.

Before proceeding we also note that a transfer of photon squeezing
to a partial spin squeezing of an ensemble of atoms 
has been suggested and demonstrated  
in \cite{polzhik1} and \cite{polzhik2}.
Here spin squeezed states are generated when an initially unexcited vapor 
absorbs non-classical light beams. In this case the transfer of non-classical 
correlations from light to atoms is however incomplete due to 
dissipation. For instance, only 50\% of spin squeezing can be achieved
by this method. Furthermore the process is irreversible.
The present paper, in contrast, suggests a general method, by 
which non-classical excitations
can be completely transfered to or from the media. In the ideal limit no 
dissipation or decoherence is present.


\section{Intracavity EIT with quantum fields}


The adiabatic transfer and storage mechanisms proposed in the present paper 
are based
on intracavity EIT \cite{OL98}. 
We therefore first review the properties 
of intracavity EIT with special emphasis
on the interaction of the combined cavity--atomic system 
with few-photon quantum fields. Recently this approach has also been applied
 to the treatment of a 
``photon blockade'' in a cavity EIT setup \cite{werner}. 
In different context,  similar 
ideas were used to describe dark states in Bose-Einstein Condensates
\cite{meystre}. 

Consider a system consisting of a single-mode cavity containing $N$ identical 
three-level atoms as shown in Fig.~1. 
Assume that one of the two optically
allowed transitions is coupled  by a cavity mode, whereas the other is 
coupled by a 
field in a coherent state. We will show later on that the
coherent field remains essentially unaffected by the interaction.
Therefore it can be represented by a time-dependent c-number  
Rabi-frequency $\Omega(t)$.  
The dynamics of this system 
is described by the interaction Hamiltonian:
\begin{equation}
H = \hbar g \sum_{i = 1}^N  \hat a\sigma_{ab}^i + 
\hbar\Omega(t) {\rm e}^{-i\nu t}
\sum_{i = 1}^N  
\sigma_{ac}^i + {\rm h.c.} .  
\label{ham}
\end{equation}
Here $\sigma_{\mu\nu}^i = |\mu\rangle_{ii}\langle \nu|$ is the 
flip operator of the $i$th atom between states $|\mu\rangle$ and $|\nu\rangle$.
$g$ is the coupling  constant between the atoms and the field mode 
(vacuum Rabi-frequency) which for simplicity is assumed 
to be equal for all atoms. 
In view of the symmetry of the coupling, it is convenient to 
introduce collective atomic operators 
$\sigma_{ab} = \sum_{i = 1}^N  \sigma_{ab}^i$ and $\sigma_{ac} = 
\sum_{i = 1}^N  \sigma_{ac}^i$. These operators couple symmetric, 
Dicke-like  states  which we denote as 
\begin{eqnarray}
|b\rangle &\equiv& |b_1...b_N\rangle,\\
|a\rangle &\equiv&  \sum_{i = 1}^N \frac{1}{\sqrt{N}} 
|b_1...a_i...b_N\rangle,\\ 
|c\rangle &\equiv& \sum_{i = 1}^N \frac{1}{\sqrt{N}}
|b_1...c_i...b_N\rangle,\\ 
|aa\rangle &\equiv& \sum_{i\ne j = 1}^N \frac{1}{\sqrt{2N(N-1)}}
|b_1...a_i...a_j....b_N\rangle,\\
|ac\rangle &\equiv& \sum_{i\ne j = 1}^N \frac{1}{\sqrt{N(N-1)}}
|b_1...a_i...c_j..b_N\rangle,\enspace{\rm etc.}
\end{eqnarray}
Quantum and classical fields cause transitions between these states
as indicated in Fig.~1.  

Under conditions of two-photon resonance, i.e.~when the energy difference 
between levels $c$ and $b$ equals the energy difference per photon of the
two fields, i.e. when 
$\omega_{cb}=\nu-\nu_c$, $\nu$ and $\nu_c$ being the frequencies
of the classical drive field and the cavity mode, the 
interaction Hamiltonian (\ref{ham}) has families of ``dark'' eigenstates 
with zero eigenvalues. These states decouple from both quantum and classical 
fields by interference. For example, the  
dark state (Fig.1b) involving at most one cavity photon corresponds to  
\begin{equation}
|D,1\rangle = -i{\Omega |b,1\rangle - g \sqrt{N} 
|c,0\rangle \over \sqrt{\Omega^2 + g^2 N} }= -i\cos\theta(t)\, |b,1\rangle
+i\sin\theta(t)\, |c,0\rangle,
\label{dark}
\end{equation}
where we have introduced the mixing angle $\theta(t)=\, {\rm arctan}\,
[g\sqrt{N}/\Omega(t)]$. 
This state has a form analogous to that of the 
usual dark state formed by a pair
of coherent classical fields. In particular, we note that in the limit 
$g\sqrt{N} \gg \Omega$ the state $|D,1\rangle$ corresponds nearly 
identically to the state $|c,0\rangle$. In this case a single-photon 
excitation is, in essence, shared among the atoms. 

Let us now discuss the principle of intracavity EIT as introduced in 
Ref. \cite{OL98}. To this end 
we include dissipation and decays into the analysis. Three
important mechanisms corresponding to such dissipation should be 
distinguished. First of all, we note that the states of the type 
given by Eq. (\ref{dark}) are immune against decay from the excited atomic
levels, as they contain no component of such states.
The dark state however is sensitive to the decay of the lower level
coherence between levels $b$ and $c$. This decay ($\gamma_{bc}$)  
sets the ultimate upper limit on the lifetime of the dark state $|D\rangle$.
Finally, there is the effect of the finite $Q$-value of the cavity.
A bare-cavity decay with a rate  $\gamma$ leads to a decay of the
dark state $|D,1\rangle$ with the effective rate
\begin{equation}
\frac{\gamma_D}{2}=\frac{\gamma}{2}\, \cos^2\theta(t)\label{gammaD}.
\end{equation}
Thus for $\cos^2\theta \ll 1$, i.e. for $g\sqrt{N} \gg \Omega$
the effect of the cavity decay is substantially reduced.
 In this limit, 
a superposition given by Eq.(\ref{dark}) contains only a very small 
($\sim \Omega/g\sqrt{N}$) component of the single-photon state 
$|b,1\rangle$.
 This increases  the lifetime of 
the combined atom-cavity system and is the essential feature 
of intracavity EIT. 

Before concluding we note another interesting property  of intracavity EIT, 
which is important for our present purposes. By changing the Rabi-frequency 
of the classical driving field $\Omega(t)$, i.e. by
varying the mixing angle $\theta(t)$, one can change the coupling of the 
cavity-dark state to the environment. In what follows we show that 
this will allow us to effectively
load the cavity system with an excitation resulting from an incoming
photon wave packet and to subsequently release this energy into a desired 
photon packet after some storage period.


\section{Manipulation of single-photon excitation by adiabatic following}


\subsection{coupling of cavity-dark state to free-field modes}


We now  discuss the problem of transferring a single-photon state
of the free field to a single-photon cavity dark state 
and vice versa.
We will show that these processes can be achieved by
{\it adiabatically rotating} the cavity dark state in a specific way.
We consider an effective one-dimensional model with a Fabry-Perot 
type cavity as shown in Fig.~2. 
The $z$-axis is parallel to the propagation of the
input and outgoing modes. $z=0$ characterizes the position
of the partially transmitting input mirror of the cavity. The other 
mirror of the cavity is assumed to be 100\% reflecting.

To model the input-output processes we introduce a continuum of 
free-space modes with field operators ${\hat b}_k$ which are coupled to the 
selected cavity mode  with a coupling constant $\kappa$. 
For simplicity we assume that the coupling constant is the same
for all relevant modes. 
This interaction is described by the 
following effective Hamiltonian  
\begin{equation}
V_{\rm cav-free} = \hbar\sum_k \kappa {\hat a}^\dagger {\hat b}_k + {\rm h.c.}.
\end{equation}
We consider an input field in  a general single-photon state
$ |\Psi_{\rm in}(t)\rangle = \sum_k\xi_k^{\rm in}(t)|1_k\rangle$
with $\xi_k^{\rm in}(t)=\xi_k^{\rm in}(t_0)\, {\rm e}^{-i\omega_k (t-t_0)}$.
Here $| 1_k \rangle$ stands for $|0,\dots,1_k,\dots,0\rangle$
and $\sum_k |\xi_k^{\rm in}|^2=1$. In what follows we describe these fields
by an envelope ``wave function'' $\Phi_{\rm in}(z,t)$ defined by:
\begin{equation}
\Phi_{\rm in}(z,t)=\sum_k \bigl\langle  0_k \bigr|\, 
{\hat b}_k\, {\rm e}^{ikz} \, 
\bigl|\Psi_{\rm in}(t)\bigr\rangle.
\end{equation}
In a continuum limit we have
$\xi_k(t)\to\xi(\omega_k,t)$ and $\sum_k \to (L/2\pi)\int {\rm d}k$
where $L$ is the quantization length. Hence
\begin{equation}
\Phi_{\rm in}(z,t)=
\frac{L}{2\pi c}\int{\rm d}\omega_k\, \xi^{\rm in}(\omega_k,t)\, {\rm e}^{ikz}.
\end{equation}
The normalization condition 
$(L/2\pi c)\int {\rm d}\omega_k |\xi^{\rm in}(\omega_k,t)|^2=1$ 
of the Fourier coefficients implies the normalization of the
input wave-function
\begin{equation}
\int\frac{{\rm d} z}{L}\, \left|\Phi_{\rm in}(z,t)\right|^2 =1.\label{Phi_z}
\end{equation}
%
%


\subsection{input-output problem}


When the single-photon wave-packet interacts with the 
combined system of cavity mode and atoms,
the general state can be written in the form:
\begin{equation}
|\Psi(t) \rangle = b(t)
\bigl|b,1, 0_k \bigr\rangle+ c(t) 
\bigl|c,0, 0_k \bigr\rangle+  a(t) |a,0, 0_k \bigr\rangle + 
\sum_{k} \xi_k(t) \bigl|b,0,1_k \bigr\rangle, 
\end{equation}
where, for example, $|b,1, 0_k \rangle$ denotes the state corresponding to 
the atomic system in the collective state $|b\rangle$, the cavity mode 
in the single-photon state
and there are no photons in the outside modes.
We now assume that the bare frequency of the cavity mode coincides with
the $a-b$ transition frequency of the atoms as well as the
carrier frequency of the input wave packet, i.e. $\nu_c=\omega_{ab}
\equiv \omega_a-\omega_b=\omega_0$. Furthermore we assume that the
classical driving field is tuned to resonance with the $a-c$ transition,
i.e. $\nu=\omega_{ac}$. This also implies that the system is in perfect 
two-photon resonance. Under these conditions we can make a transformation
into a frame rotating with optical frequencies. 
The following equations of motion describe the evolution of the 
slowly-varying state amplitudes:
\begin{eqnarray}  
{\dot a}(t) &=&  -\frac{\gamma_a}{2}\, a(t) 
-i g\sqrt{N} b(t) -i\Omega  c(t),\label{adot}\\
{\dot b}(t) &=&  -ig\sqrt{N} a(t)
-i\kappa\sum_k\xi_k(t),\label{bdot}\\
{\dot c}(t) &=& -\frac{\gamma_c}{2}\, c(t)
-i\Omega  a(t),\label{cdot}\\
{\dot\xi}_k(t) &=&-i\Delta_k\xi_k(t) -i\kappa b(t),\label{xidot}
\end{eqnarray}
where $\Delta_k=\omega_k-\omega_0= kc -\omega_0$ is the detuning of the
free-field modes from the cavity resonance, and  $\omega_0=\nu_c=\omega_{ab}$.
In order to model the decay processes such as 
spontaneous emission and the finite lifetime
of the state $c$ (and ultimately the dark state) we use an open 
system approach and  introduce
decay rates $\gamma_a$ and $\gamma_c$ out of the system.  

We note the enhancement of the coupling of atoms with the cavity mode
by a factor $\sqrt{N}$ due to collective effects.
At the same time, however,  no such enhancement of the decay rates
$\gamma_a$ and $\gamma_c$ takes place as the decays affect 
the atoms individually. In the following we assume
that $\gamma_c$ is sufficiently small. In this case it can be ignored
during the time required for the input and the output processes.
$\gamma_c$ will be taken into account however for  the
storage time interval.

To describe the adiabatic transfer we proceed by introducing a 
basis of dark and bright states, $|D\rangle$
and $|B\rangle$ \cite{Cohen-Tannoudji}:
\begin{eqnarray}
|D\rangle &=& -i\cos\theta(t)\, \bigl|b,1, 0_k \bigr\rangle 
                +i\sin\theta(t)\, \bigl|c,0, 0_k \bigr\rangle,\\
|B\rangle &=& \sin\theta(t)\, \bigl|b,1, 0_k \bigr\rangle +
\cos\theta(t)\, \bigl|c,0, 0_k \bigr\rangle,
\end{eqnarray}
where $\tan\theta(t)=g\sqrt{N}/\Omega(t)$. The evolution equations
can be re-written in terms of corresponding probability amplitudes as 
\begin{eqnarray}  
{\dot{  a}}(t) &=& 
-\frac{\gamma_a}{2}\, {  a}(t) 
-i \Omega_0(t) \, {  B}(t),\\
{\dot{  B}}(t) &=& -i\dot\theta(t)\, D(t) -i\Omega_0
\, {  a}(t) -i\kappa\sin\theta(t)\sum_k{ \xi}_k(t),\\
{\dot{  D}}(t) &=& -i\dot\theta(t)\, B(t) +\kappa\cos\theta(t)
\sum_k{ \xi}_k(t),\\
{\dot{  \xi}}_k(t) &=&-i\Delta_k{ \xi}_k(t) 
-i\kappa\sin\theta(t){  B}(t)-\kappa\cos\theta(t) {  D}(t).
\end{eqnarray}
Here $\Omega_0(t)=\sqrt{g^2 N+\Omega^2(t)}$, and the terms proportional
to $\dot\theta$ describe the non-adiabatic coupling between the bright and
dark state. 
We now adiabatically eliminate the excited state, which is possible if the
characteristic time $T$ of the process is sufficiently large compared to 
the radiative lifetime of the excited state ($\gamma_a T\gg 1$). 
In a second step we adiabatically
eliminate the bright-state amplitude and disregard
non-adiabatic corrections. The conditions under which such an
elimination is justified will be discussed later.
We finally arrive at
\begin{eqnarray}
{\dot{  D}}(t) &=& \kappa\cos\theta(t)\sum_k{ \xi}_k(t),
\label{D1}\\
{\dot{  \xi}}_k(t) &=& -i\Delta_k\, { \xi}_k(t) -\kappa
\cos\theta(t)\, {  D}(t).\label{xi}
\end{eqnarray}
One immediately recognizes from these equations, that
the total probability of finding
the system in a free-field single photon state or in the cavity-dark state
is conserved
\begin{equation}
\frac{{\rm d}} {{\rm d }t} \left(\left|D(t)\right|^2+\sum_k 
\left|\xi_k(t)\right|^2\right) = 0.
\end{equation}
Thus under adiabatic conditions 
there is only an exchange of probability between the free-field
states and the cavity dark state.

Formally integrating Eq.(\ref{xi}) leads to
\begin{eqnarray}
{ \xi}(\omega_k,t) &=& { \xi}^{\rm in}(\omega_k,t_0)\, 
{\rm e}^{-i\Delta_k(t-t_0)}
-\kappa\int_{t_0}^t\!\! {\rm d}\tau\, \cos\theta(\tau)
{  D}(\tau)\, {\rm e}^{-i\Delta_k(t-\tau)}\label{xi_int}
\end{eqnarray}
and therefore
\begin{eqnarray}
\dot{  D}(t) &=& \frac{\kappa L}{2\pi c}
\, \cos\theta \int{\rm d}\omega_k\, { \xi}^{\rm in}(\omega_k,t_0)\,
{\rm e}^{-i\Delta_k(t-t_0)}\nonumber\\
&&-\kappa^2 \cos\theta(t)\int_{t_0}^t\!\!{\rm d}\tau
\cos\theta(\tau){  D}(\tau)\frac{L}{2\pi c}
\int{\rm d}\omega_k {\rm e}^{-i\Delta_k(t-\tau)}.
\end{eqnarray}
In the first term we can identify the wave function of the
input photon at $z=0$. 
Furthermore in the Markov-limit 
$\int{\rm d}\omega_k\, {\rm e}^{-i\Delta_k(t-\tau)}
\to 2\pi\delta(t-\tau)$. Thus we find
\begin{eqnarray}
\dot{  D}(t) &=& \sqrt{\gamma{c \over L}}\cos\theta(t)
{  \Phi}_{\rm in}(0,t)
-\frac{\gamma}{2}\cos^2\theta(t) {  D}(t)\label{DE_D1}
\end{eqnarray}
where we have introduced the empty-cavity decay rate
$\gamma=\kappa^2 L/c$.
If $t_0$ is a time sufficiently before any excitation of the
cavity system takes place, 
i.e. if $\Phi_{\rm in}(0,t)=0$ for all $t\le t_0$, 
 the solution of (\ref{DE_D1})
can be written as
\begin{equation}
{  D}(t)=\sqrt{\gamma {c \over L}}\int_{t_0}^t\!\!{\rm d}\tau\,
\cos\theta(\tau)\, { \Phi}_{\rm in}(0,\tau)
\,\exp\left\{-\frac{\gamma}{2}\int_\tau^t\!\!{\rm d}\tau^\prime
\, \cos^2\theta(\tau^\prime)\right\}.\label{Dres}
\end{equation}
Substituting Eq. (\ref{Dres}) into Eq. (\ref{xi_int}) leads to the
 input-output relation
\begin{eqnarray}
\Phi_{\rm out}(0,t)&=&{ \Phi}_{\rm in}(0,t)\nonumber\\ 
&&-\gamma\, \cos\theta(t)\int_{t_0}^t\!\!{\rm d}\tau\, \cos\theta(\tau)
\,\Phi_{\rm in}(0,\tau)\, \exp\left\{-\frac{\gamma}{2}
\int_\tau^t\!\!{\rm d}\tau^\prime
\, \cos^2\theta(\tau^\prime)\right\}.
\label{in-out}
\end{eqnarray}

Before proceeding let us consider the conditions for
the adiabatic elimination of the bright-state amplitude. For this we
substitute the formal integral (\ref{xi_int}) into Eqs.(\ref{adot}-
\ref{cdot}) and take the Markov-limit. We then find that
adiabatic following occurs when
\begin{equation}
\Omega_0^2\gg\gamma\gamma_a,\quad\Omega_0^2\gg\frac{\gamma_a}{T},
\quad \Omega_0^2\gg\sqrt{\frac{\gamma}{T}}\, \gamma_a.
\label{adiabat_cond}
\end{equation}
%
%
We note that these conditions also enshure that spontaneous Raman
scattering in other than the cavity mode are negligible.
Since the characteristic input-pulse length and thus the characteristic
times $T$ have to be larger or equal to the cavity decay time $\gamma^{-1}$,
the first condition is the most stringent one. 

It is important to note that in order to  
ensure adiabaticity it is sufficient that
\begin{equation}
g^2 N \gg \gamma\gamma_a.
\end{equation}
This condition should be contrasted to the corresponding condition
of adiabatic transfer with a single atom. The single-atom case
requires a strong-coupling regime corresponding (at least) to 
$g^2 \ge \gamma\gamma_a$ \cite{Parkins93}. The latter is  
very difficult to realize experimentally.

Let us now discuss the implications of Eqs.(\ref{Dres}) and (\ref{in-out}).
If $\cos\theta$ is constant in time, the atoms simply cause 
a change of the cavity decay rate, according to $\gamma \to 
\gamma\cos^2\theta$,
Eq.(\ref{gammaD}). Hence, by increasing the atom density and therefore
decreasing $\cos\theta$, the effective lifetime of the cavity mode 
can be increased. This is however of no help 
if we are interested
in ``storing'' a photon wave packet. When the effective $Q$-value of the
cavity is increased, the resonances of the combined atom-cavity system 
become extremely narrow and  the outgoing wave packet is smeared out 
in time. Furthermore there is an
increasing component corresponding to the 
input field directly reflected from the input mirror. Clearly 
the transfer of photons from an
input pulse into the cavity deteriorates significantly when the
pulse length becomes shorter than the effective cavity decay time.
This is illustrated in Fig. 3, where we 
have shown the input and output wave functions for different values of the
effective cavity decay. The input wave function is a hyperbolic secant pulse.
 
We now describe a method which allows one to capture and to subsequently 
release a 
single-photon state of the light field. In order to achieve this, we 
utilize techniques of adiabatic transfer \cite{STIRAP}.  
To motivate the analysis carried 
out below we note that the state $|D,1\rangle$, Eq. (\ref{dark}) couples
to the free-field light modes only due to the admixture of the state 
$|b,1\rangle$. 
As can be seen from Eq.(\ref{D1}) the coupling of the dark state
to the free-field light modes depends on the cosine of the
mixing angle $\theta$.
When the Rabi-frequency of the classical field
$\Omega$ is large, $\cos\theta$ is large and there is a strong coupling 
between cavity-dark state and free field. In this
case the free-field photons can leak in an out of  the cavity. However, when 
$\Omega$ is small this leakage is effectively stopped. Therefore, by first 
accumulating the field in a cavity mode and then adiabatically 
switching off the driving field, an initial free-space wave packet 
can be stored in a long-lived atom-like dark state. The latter can 
be released by simply reversing the process, i.e. by an
(adiabatic) increase of the Rabi-frequency of the driving field. 
These two processes will now be discussed in detail.


\subsection{optimization of input: quantum impedance matching}


In this section we show how to optimize the time 
dependence of $\cos\theta(t)$ such that the dark-state amplitude 
will asymptotically come close to
unity. It is clear at hand that this is only possible for a
bandwidth of the incoming wave function which is less
or at most equal to the bare-cavity bandwidth, i.e. for a wave-packet which
is longer than the bare-cavity decay time. Also the time when the 
adiabatic transfer starts must coincide with the  arrival 
time of the photon wave-packet. 

In order to achieve a maximum
transfer of free-field photons into cavity photons, the 
outgoing field components should be minimized. This can be done for example  
by using the destructive
interference of the directly reflected
and the circulating components. A necessary condition for 
complete destructive interference 
can be obtained by differentiating the input-output relation
Eq. (\ref{in-out}) and setting $\Phi_{\rm out}={\dot\Phi}_{\rm out}=0$.
This yields
\begin{eqnarray} 
-\frac{{\rm d}}{{\rm d} t}  \, {\rm ln}\, \cos\theta(t)
+\frac{{\rm d}}{{\rm d} t} \, {\rm ln}\, \Phi_{\rm in}(t)
=\frac{\gamma}{2}\cos^2\theta(t).\label{impedance}
\end{eqnarray}
This equation has a simple physical interpretation. 
The first term on the l.h.s. is the amplitude loss rate of the
photon field inside the cavity. 
When the rotation angle $\theta$ is increased  by decreasing the
Rabi-frequency of the classical driving field, the atoms will
absorb photons from the cavity mode to re-establish the dark state
by a Raman transition from $|b\rangle$ to $|c\rangle$. 

The term on the
right-hand side is the effective amplitude decay rate due to cavity losses.
Thus if $\Phi_{\rm in}$ would be constant, Eq.(\ref{impedance})
constitutes, what in classical systems 
is known as {\it impedance matching condition}\cite{Siegmann}. 
Under conditions of impedance
matching, there is complete destructive interference of the 
directly reflected part of the incoming wave and the circulating field
leaking out through the input mirror. 
The classical impedance-matching
condition needs to be modified when the input field is time-dependent,
as the circulating field ``sees'' a slightly changed
input field after a cavity-round trip.
This then leads to the  second term on the l.h.s. of  Eq.(\ref{impedance}).
An intuitive derivation of this term as well as a simple physical explanation
of the quantum impedance matching condition is given in the Appendix.

We now illustrate the remarkable performance of the 
adiabatic transfer mechanism
under conditions of quantum impedance matching. 
Since Eq.(\ref{impedance}) depends explicitly 
on the pulse shape, let us specify a particular 
form of the input pulse. Consider, for example,  the
case of a normalized hyperbolic secant input pulse
\begin{equation}
\Phi_1(t)=\Phi_{\rm in}^{(1)}(z=0,t)=
\sqrt{{L \over c T}}\, {\rm sech}\, \left[{2t\over T} \right].
\end{equation}
%
%
The quantum impedance matching condition leads to the nonlinear 
first-order differential equation 
\begin{equation}
\frac{{\rm d}}{{\rm d} t} \cos\theta(t) 
+\frac{\gamma}{2}\cos^3\theta(t)
+\frac{2}{T} \tanh\bigl[2 t/T\bigr] \cos\theta(t)=0.\label{DE_cos}
\end{equation}
Eq.(\ref{DE_cos}) can be solved analytically and we are looking for
solutions with the asymptotic behavior $\cos\theta \rightarrow 0$
for $t\to\infty$ . One of such solutions corresponds to 
\begin{equation}
\cos\theta(t)=\sqrt{\frac{2}{\gamma T}}\, \frac{\, {\rm sech}\,[2t/T]}
{\sqrt{1+
\tanh[2t/T]}}.
\label{cos}
\end{equation}
The specific form of the mixing angle 
given by the above equation can be achieved, 
provided that the single-photon pulse is 
long enough ($\gamma T \ge 4$),  by changing the Rabi-frequency 
of the classical driving field according to: 
\begin{equation}
\Omega(t) = g \sqrt{N} \frac{{\rm sech}(2t/T)}{\sqrt{[1+{\rm tanh}(2t/T)]
[{\rm tanh}(2t/T) + \gamma T/2 -1]
}}.
\end{equation}
With this choice for the driving field 
one finds 
that the dark-state population 
corresponding to an input field $\Phi_1$ evolves according to: 
\begin{equation}
|D(t)|^2 =  { 1+ {\rm tanh}[2t/T] \over 2}. 
\end{equation}
Clearly the population of the dark state 
approaches unity as $t\to \infty$. This is illustrated in Fig.~4.

An obvious {\it disadvantage} of the quantum impedance matching condition
 Eq.(\ref{impedance}) is 
its explicit dependence on the shape of the input pulses $\Phi_{\rm in}$. 
We will now show that the asymptotic 
population of the dark state is, in fact,  not very sensitive 
to the actual shape. To illustrate this,
we have plotted in Fig.~4
the time dependence of the dark-state
population for a Gaussian input field
\begin{equation}
\Phi_2(t)=\Phi_{\rm in}^{(2)}(z=0,t)=
\sqrt{{L \over  cT}}\, \left(\frac{2}{\pi}\right)^{1/4}
\exp\Bigl\{-{t^2 \over T^2}\Bigr\}
\end{equation}
as well as  for a hyper-Gaussian wave function 
\begin{equation}
\Phi_3(t)=\Phi_{\rm in}^{(3)}(z=0,t)= \sqrt{{L \over cT}}\,
\left(\frac{\Gamma\left[\frac{5}{4}\right]}{2^{3/4}}\right)^{1/2}\, 
\exp\Bigl\{-{t^4 \over T^4}
\Bigr\}.
\end{equation}
With these initial pulses we use the 
``incorrect'' mixing angle, Eq.(\ref{cos}), 
chosen to optimize the input for a hyperbolic secant pulse. By numerically
integrating the equations of motion, we find 
the asymptotic values of the dark-state amplitudes are in these
cases $D\to 0.9942$ and  $D\to 0.9778$ respectively. This indicates
that there is only a modest dependence upon the actual shape of the input
pulse for a given function $\cos\theta(t)$. 

It should be noted that an exact timing of the arrival time is essential. 
A small delay $\delta t$ 
in the arrival time of the pulses leads to a decrease of the asymptotic 
amplitude of the dark state proportional to $\delta t^2$.

In the above discussion we have assumed that the external
control field is at all time in a coherent state and have 
represented it by its coherent amplitude $\Omega(t)$. This
assumption is only valid if the drive field remains
unaffected by the interaction with the ensemble of atoms
even when its intensity is turned to zero. This is however the case
here, since although $\Omega(t)\to 0$, the ratio of $\Omega(t)$ 
to the effective Rabi-frequency of the field mode 
$g\sqrt{\langle n(t)\rangle}$ is always much larger than unity. In 
fact in the case of impedance matching one finds the
asymptotic behavior $\Omega(t)/g\sqrt{\langle n(t)\rangle}\to \sqrt{N}$.


\subsection{output}


In order to release the stored photon into free-field photons  at some
later time $t_1$, one can simply reverse the adiabatic rotation of the
mixing angle. The resulting wave-packet will not necessarily have the
same pulse form as the original one. The latter aspect is not essential
for the purposes of quantum information processing. It is however 
important that
the output wave-packet is generated in a well defined way and corresponds,
in the ideal limit, to a single-photon Fock state. 

For a time $t_1$ large enough, such that $\Phi_{\rm in}(0,t)=0$
for all $t>t_1$, and for $\cos\theta(t_1)=0$ we find from 
the input-output relation
\begin{eqnarray}
{ \Phi}_{\rm out}(t)
&=&-\sqrt{\frac{\gamma L}{c}}\,
D(t_1)\cos\theta(t)\,\exp\left\{-\frac{\gamma}{2}
\int_{t_1}^t\!\!{\rm d}\tau\, \cos^2\theta(\tau)\right\}.
\end{eqnarray}
Thus the shape of the output wave-packet is entirely determined by
the function 
$\cos\theta(t)$. For the time-reversal of Eq.(\ref{cos}) a hyperbolic
secant output pulse is generated. This is illustrated in Fig.~5.
If the dark-state decay during the unloading period is again neglected,
the amplitude of the output wave function depends on the dark state
amplitude at the release time only. One easily verifies that the total
number of photons in $\Phi_{\rm out}$ is given by
\begin{eqnarray}
\frac{c}{L}\int_{t_1}^\infty\!\! {\rm d}t \left|{ \Phi}_{\rm out}(t)\right|^2
=\left|{ D}(t_1)\right|^2.
\end{eqnarray}

The ultimate fidelity of the storage is determined by the decay of 
the collective dark state during
the storage time. Under reasonable conditions the dark-state 
decay can be neglected during the loading and unloading periods. Hence  
we only need to determine how $D(t_1)$ (at the time of the release) differs 
from $D(t_0)$ (at the time of arrival), where
$t_1-t_0$ is the storage time. If we take into account a decay
out of the atomic level $|c\rangle$  with
a single-atom decay rate $\gamma_c$, we find the simple result
\begin{equation}
D(t_1)=D(t_0)\, \exp\left\{-\frac{\gamma_c}{2} (t_1-t_0)\right\}.
\end{equation}
It is worth noting that the decay of the collective dark state
is identical to  the {\it single-atom} decay. This may seem as a surprise
on first glance, since the coupling strength to the
cavity mode is enhanced by a factor 
$\sqrt{N}$.
One should bear in mind however that 
the decay affects only those atoms which are in state $c$ and that
in the collective dark state
each atom has only a 
probability of $1/N$ to be in that state.

 
\section{Transfer and storage of non-classical superposition states}


A convenient way of encoding quantum information in photons is to use
the analogy between spin-1/2 systems and polarization states. 
We therefore include polarization 
of the quantum field and study the interaction of superpositions of 
polarization states with the 
intracavity EIT system. 

Let us consider a quantum field consisting of a right ($\sigma_+$) and left 
($\sigma_-$) circularly polarized components interacting 
 with a multi-state system shown in 
Fig. 6a. The system is driven by a classical driving field
of different polarization and frequency characterized by the
time-dependent Rabi-frequency $\Omega$.

We assume that initially all 
population is in the lower state $|b\rangle$ coupled by 
both $\sigma_+$ and $\sigma_-$ components. We consider here the interaction 
of such atomic ensemble with 
a single photon 
wave-packets of the type 
\begin{eqnarray} 
|\Psi_{\rm in}(t)\rangle = 
\sum_k\xi_{+k}^{\rm in}(t) |1_{+k}\rangle |0_{-k}\rangle 
+\sum_k\xi_{-k}^{\rm in}(t) |0_{+k}\rangle |1_{-k}\rangle.
\label{stateff}
\end{eqnarray} 
$|\Psi_{\rm in}\rangle$ is an eigenstate of the photon number operator
$\hat n \equiv\hat n_+ +\hat n_-$ with eigenvalue unity, i.e.
$\sum_k\Bigl(|\xi_{+k}|^2+|\xi_{-k}|^2\Bigr)=1$.
Since polarization states are distinguishable 
one immediately recognizes that the interaction of atoms and cavity 
separates into two
families of states, which do not couple to each other. This
is illustrated in Fig. 6b. Thus the state vector of the
interacting system can be written as
\begin{eqnarray}
|\Psi(t)\rangle &=& |\Psi_+(t)\rangle\, |0_-\rangle 
+|\Psi_-(t)\rangle\, |0_+\rangle,\\
|\Psi_+(t)\rangle &=& 
 b_+(t)
\bigl|b,1_+, 0_{+k} \bigr\rangle+ c_+(t) 
\bigl|c_+,0_+, 0_{+k} \bigr\rangle+  a_+(t) |a_+,0_+, 0_{+k} \bigr\rangle 
\nonumber\\
&& + \sum_{k} \xi_{+k}(t) \bigl|b,0_+,1_{+k} \bigr\rangle, \\
|\Psi_-(t)\rangle &=& 
 b_-(t)
\bigl|b,1_-, 0_{-k} \bigr\rangle+ c_-(t) 
\bigl|c_-,0_-, 0_{-k} \bigr\rangle+  a_-(t) |a_-,0_-, 0_{-k} \bigr\rangle 
\nonumber\\
&&+ \sum_{k} \xi_{-k}(t) \bigl|b,0_-,1_{-k} \bigr\rangle. 
\end{eqnarray}
The equations of motion for the state amplitudes separate into 
two sets, identical to Eqs.(\ref{adot}-\ref{xidot}). We thus can proceed
in exactly the same way as in the previous section.
In particular we  introduce the dark states  
\begin{eqnarray} 
|D_+ \rangle &=& {\Omega |b,1_+,0_-\rangle - g \sqrt{N} 
|c_+,0_+,0_-\rangle \over \sqrt{\Omega^2 + g^2 N} }, \\
|D_- \rangle &=& {\Omega |b,0_+,1_-\rangle - g \sqrt{N} 
|c_+,0_+,0_-\rangle \over \sqrt{\Omega^2 + g^2 N} },
\end{eqnarray} 
where $0_\pm$ and $1_\pm$  denote the cavity-mode excitation and we have 
dropped the free-field component for simplicity. 
In the adiabatic limit the total number of excitations in both
sub-systems is constant, i.e. 
\begin{eqnarray} 
\frac{{\rm d}} {{\rm d }t} \left(\left|D_+(t)\right|^2+\sum_{k} 
\left|\xi_{+k}(t)\right|^2\right) &=& 0,\\
\frac{{\rm d}} {{\rm d }t} \left(\left|D_-(t)\right|^2+\sum_{k} 
\left|\xi_{-k}(t)\right|^2\right) &=& 0.
\end{eqnarray}
Let us now consider the case when the initial wave packet is in a  coherent
superposition of two polarization states with identical envelopes, i.e.
\begin{eqnarray} 
\xi^{in}_{+k}(t) = \alpha\, \xi^{in}_{k}(t), \quad \quad 
\xi^{in}_{-k}(t) = \beta\, \xi^{in}_{k}(t).
\end{eqnarray}
In this case the adiabatic following technique described above can be 
performed for both polarizations  in parallel yielding, apart from overall
constants, an identical evolution of the dark state amplitudes 
$|D_{\pm}\rangle$. The general 
state of a free field (\ref{stateff}) can therefore be transfered 
back and forth to a collective atomic state
\begin{eqnarray} 
|\Psi_{\rm in}\rangle \longleftrightarrow  \Bigl[
\alpha |c_+\rangle + \beta |c_-\rangle\Bigr] |0_+,0_-\rangle. 
\end{eqnarray} 
We note in particular that the relative phase between the left- and
right-circularly polarized input wave packets is mapped onto the relative
phase between the collective atomic states $|c_+\rangle$ and 
$|c_-\rangle$. Hence quantum mechanical superposition states can be 
``stored'' in collective atomic excitations. 

Before concluding we remark that much more general field 
states  can be transfered onto the atoms. Consider for instance an 
entangled state composed of two single-photon
states of different polarization. 
Of particular interest are maximally entangled superpositions such as 
$\sim |0_+,0_-\rangle +|1_+,1_-\rangle$. An input state of this form
 contains a zero- and a two-photon component. 
Using the adiabatic
techniques of the present paper it is also possible to transfer
states of this kind onto collective atomic states. The theoretical 
description of the interaction is however more involved, as it requires
invoking higher-order  dark states. In particular, for mapping such
entangled two-photon states 
onto atoms, two additional dark states play an important role:
\begin{eqnarray} 
|D_0 \rangle &=& |b,0_+,0_-\rangle, \\
|D_2 \rangle &=& {\Omega^2 |b,1_+,1_-\rangle - g \sqrt{N} \Omega( 
|c_+,0_+,1_-\rangle  + 
|c_-,1_+,0_-\rangle) + g^2 \sqrt{N(N-1)} |c_+,c_-,0_+,0_-\rangle  
\over \sqrt{\Omega^4 +2 g^2 N\Omega^2 
 + g^4 N(N-1)}}.
\end{eqnarray} 
It is obvious at the intuitive level that, in  the  ideal limit, 
an adiabatic transfer  will yield 
atomic states of the type $\sim (|b\rangle +
|c_{+},c_{-}\rangle) |0_{+},0_{-}\rangle $.
At the same time we note that due to a different functional
form of the doubly excited dark state, and due to a cross-coupling between
different channels of excitation, the conditions for generating  such 
states can be somewhat different from those described in previous sections.  
The specific conditions 
as well as applications to quantum information processing will 
be discussed in detail elsewhere.


\section{Summary}


In conclusion we suggested a new technique for mapping 
quantum states of the radiation field onto collective
atomic excitations. Our approach utilizes  intracavity electromagnetically 
induced transparency and therefore does not require the usual 
strong-coupling
condition of cavity QED. By adiabatically rotating the dark state(s) of
a system consisting of a large number of multi-level atoms
interacting with a single cavity mode, quantum impedance matching
of this cavity  can be achieved for an input single-photon wave-packet.
In this case the quantum state of the radiation field  can be transferred
with nearly 100\% efficiency to a non-decaying, meta-stable
state of the atoms. The quantum states of the field can therefore be ``stored''
in long-lived atomic superpositions. Reversing the adiabatic rotation
the stored state can be transformed back into a well defined
output wave-packet.

In addition to rather direct applications for quantum memory registers,
extension of these ideas 
to  quantum networks and entanglement distribution are obvious.
If the input photon wave-packet of the system is entangled with some
other system, this entanglement is transferred to the collective atomic
state. The storage mechanism also allows to reshape the output
wavepacket with respect to the input in an (almost) arbitrary way. 
Furthermore  applications of these ideas to
elementary logic gates are likely.
We therefore anticipate important applications
in different areas of quantum information processing such as 
quantum communication and quantum computing. 

The authors thank Marlan Scully, Steve Harris, Eugene Polzik
and Atac Imamo\u glu for 
many useful discussions resulting in the present paper. We are also 
grateful to Wolfgang Schleich for putting together this special issue
and his encouragement resulting in the completion of 
this work. 
This work was supported by the National Science Foundation.  

\


\section*{Appendix}


The impedance matching condition (\ref{impedance}) can be given a
simple physical explanation. For this we consider the Fabry-Perot
cavity as shown in Fig. 2. The lossless input mirror has an amplitude 
reflectivity and transmission of $R$ and $T$, satisfying 
the usual relations $R^* T
+R T^*=0$ and
$|R|^2+|T|^2=1$. Without loss of generality we set $R^*=R$ and $T=i|T|$.
Input and output field strength and the circulating
field component are denoted by $E_{\rm in}$, $E_{\rm out}$ and $E_c$.

If the carrier frequency of the input field coincides with the cavity
resonance one has the following relations between the three field components
\begin{eqnarray}
E_c(t) &=& T E_{\rm in}(t) + R\zeta E_c(t-\tau_c),\label{Ec}\\
E_{\rm out}(t) &=& T\zeta E_c(t-\tau_c) + R E_{\rm in}(t).\label{Eout}
\end{eqnarray}
$\zeta$ denotes the amplitude losses in a single round-trip and we have
denoted the round-trip time as $\tau_c$.
Substituting (\ref{Ec}) into (\ref{Eout}) yields
\begin{equation}
E_{\rm out}(t)= R E_{\rm in}(t) +\frac{T}{R}\bigl[E_c(t)- 
T E_{\rm in}(t)
\bigr]=\frac{1}{R}  E_{\rm in}(t)+\frac{T}{R} E_c(t).
\label{Eout2}
\end{equation}

The resonator set-up is called impedance matched, if the first and second
term in Eq.(\ref{Eout2}) interfere destructively. To find a condition for such
a destructive interference, we have to determine the circulating field
in terms of the input field.
Since the round-trip time $\tau_c$ is short compared to the characteristic time
of changes in the input field, we may set $E_c(t-\tau_c)
\approx E_c(t)- \tau_c \dot E_c(t)$. We do  
keep the first time derivative here,
as it will lead to a modification of the impedance matching condition for
a time-dependent input field. We thus obtain from (\ref{Ec}) 
the differential equation
\begin{equation}
\dot E_c(t) =-\eta \, E_c(t) + \frac{T}{R \zeta 
\tau_c} E_{\rm in}(t)
\label{DE_circ}
\end{equation}
where $\eta = (1-R \zeta)/( R\zeta \tau_c)$.
Eq. (\ref{DE_circ}) has the simple solution
\begin{equation}
E_c(t) = \frac{T}{R\zeta \tau_c} \int_0^\infty \!\! {\rm d}\tau
E_{\rm in}(t-\tau) \, {\rm e}^{-\eta\tau}.
\end{equation}
For small internal losses
and a reflectivity of the input mirror near unity we have
$R\approx 1-\gamma \tau_0/2$, $T^2=R^2-1\approx -\gamma \tau_0$ and 
$\zeta \approx 1 - \gamma_{\rm int} \tau_c/2$. Here $\gamma$ is the
empty-cavity decay rate, $\tau_0$ is the empty-cavity round-trip
time, and we have introduced the effective decay rate of the circulating
field due to internal losses $\gamma_{\rm int}$.
In this limit $\eta\approx (\gamma/2) (\tau_0/\tau_c) +\gamma_{\rm int}/2$.
Thus we eventually obtain for the output field
\begin{eqnarray}
E_{\rm out}(t) &=& \frac{1}{R} E_{\rm in}(t) - \frac{1}{R}\gamma
\frac{\tau_0}{\tau_c}
\int_0^\infty \!\! {\rm d}\tau\,
E_{\rm in}(t-\tau) \, {\rm e}^{-\eta\tau}.
\end{eqnarray}
Setting $E_{\rm out}=0$, multiplying with $R$, 
and differentiating yields
\begin{eqnarray}
0={\dot E}_{\rm in} - \gamma \frac{\tau_0}{\tau_c} E_{\rm in}
+\eta E_{\rm in},
\end{eqnarray}
which can be brought into the form
\begin{eqnarray}
\frac{\gamma_{\rm int}}{2} + \frac{{\rm d}}{{\rm d} t} \, {\rm ln}\, 
E_{\rm in}(t) = \frac{\gamma}{2}\frac{\tau_0}{\tau_c}.\label{impedance_app}
\end{eqnarray}
This is the generalized impedance matching condition for a single-sided
Fabry-Perot cavity with internal losses ($\gamma_{\rm int}$),
a round-trip time $\tau_c$, and a time-dependent input field.
We will now show that for the system discussed in the present paper
$\tau_0/\tau_c=\cos^2\theta(t)$ and $\gamma_{\rm int} =- 2\frac{{\rm d}}
{{\rm d} t}\, {\rm ln} \, \cos\theta(t)$.

In order to determine the round-trip time we note, that the large linear
dispersion of the EIT medium in our system leads to a strong
group delay. The group velocity
of a weakly excited, propagating field mode interacting with $N$ 
$\Lambda$-type atoms is given by 
\begin{equation}
v_{\rm gr}= \frac{c}{1+\displaystyle{
\frac{g^2 N}{\Omega^2(t)}}} = \frac{c}{1+\tan^2\theta(t)}
=c\,\cos^2\theta(t),
\end{equation}
where $\Omega(t)$ is the Rabi-frequency of the classical driving field
and $g$ describes the atom-field coupling strength. Thus 
\begin{equation}
\frac{\tau_0}{\tau_c} =\cos^2\theta(t).
\end{equation}

In order to determine the internal photon losses in the system, we
consider the equation of motion for the probability to find a single photon
inside the cavity, which is identical to the probability to find the system in
state $|b,1,0_k\rangle$. Under adiabatic conditions, the system is
always in the dark state $|D\rangle$, thus the 1-photon probability
reads 
\begin{equation}
p_1(t) = \Bigl|\bigl\langle b,1,0_k\bigr|\bigl.D(t)\bigr\rangle 
\Bigr|^2 = 
\cos^2\theta(t).
\end{equation}
Differentiating this expression with respect to time yields
\begin{equation}
\gamma_{\rm int}\equiv -\frac{{\rm d}}{{\rm d} t} 
\, {\rm ln}\, p_1(t) = -2\frac{{\rm d}}{{\rm d} t} 
\, {\rm ln}\, \cos\theta(t)
\end{equation}
With this, Eq.(\ref{impedance_app}) goes over into 
the quantum impedance condition Eq.(\ref{impedance}).


\def\etal{\textit{et al.}}


\

\begin{figure}[ht]
\vspace*{2 ex}
 \centerline{\epsfig{file=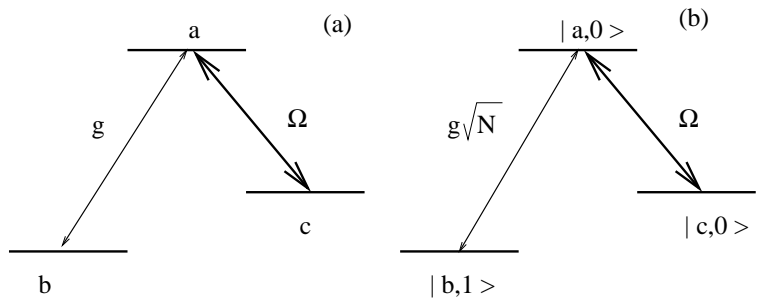,width=10.0cm}}
 \vspace*{2ex}
 \caption{(a) Three-level atoms interacting with quantum field and driven 
by classical field with Rabi frequency $\Omega(t)$. $g$ is 
the coupling constant between quantum field and atoms. 
(b) Interaction of singly 
excited mode with $N$ 3-level atoms in the basis of collective states.
\label{fig1.fig}
}
\end{figure}


%
%
\begin{figure}[ht]
 \centerline{\epsfig{file=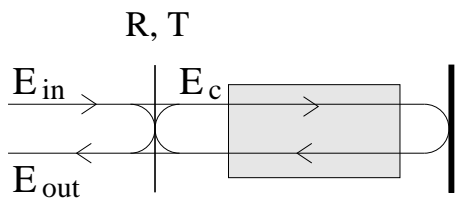,width=6.0cm}}
 \vspace*{2ex}
 \caption{Cavity set-up. $R$ and $T$ are amplitude reflectivity and 
transmittivity of input mirror. $E_{\rm in}$, $E_{\rm out}$ and $E_c$
denote input, output and circulating field components.
 	\label{fig2.fig}
}
\end{figure}


\begin{figure}[ht]
 \centerline{\epsfig{file=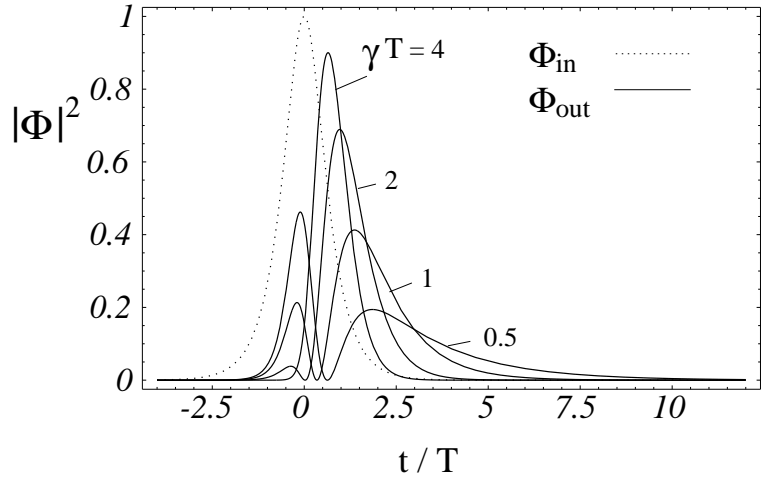,width=10.0cm}}
 \vspace*{2ex}
 \caption{Shape of input and output single-photon wave functions
of a Fabry-Perot-type resonator for different cavity decay rates. 
Decreasing of cavity width leads
to delocalized output wave function and increasing 
component reflected at $t=0$. $T$ characterizes the time unit.
 \label{fig3.fig}
}
\end{figure}


\begin{figure}[ht]
 \centerline{\epsfig{file=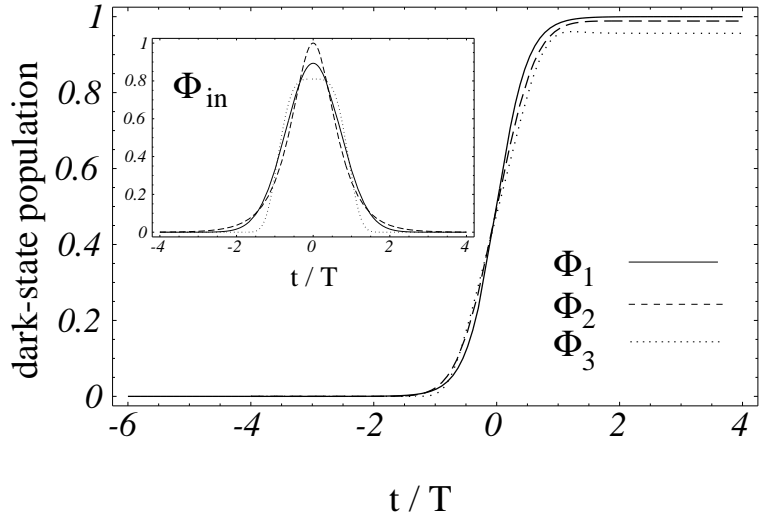,width=10.0cm}}
 \vspace*{2ex}
 \caption{Population of dark state $|D(t)|^2$ for
hyperbolic secant ($\Phi_1$), Gaussian ($\Phi_2$), and
hyper-Gaussian ($\Phi_3$) input. $\cos\theta(t)$ is optimized
for quantum impedance matching of $\Phi_1$. 
$\gamma T =4$. Shape of input wave functions shown in inset.
$T$ characterizes the time unit.
\label{fig4.fig}
}
\end{figure}


\begin{figure}[ht]
 \centerline{\epsfig{file=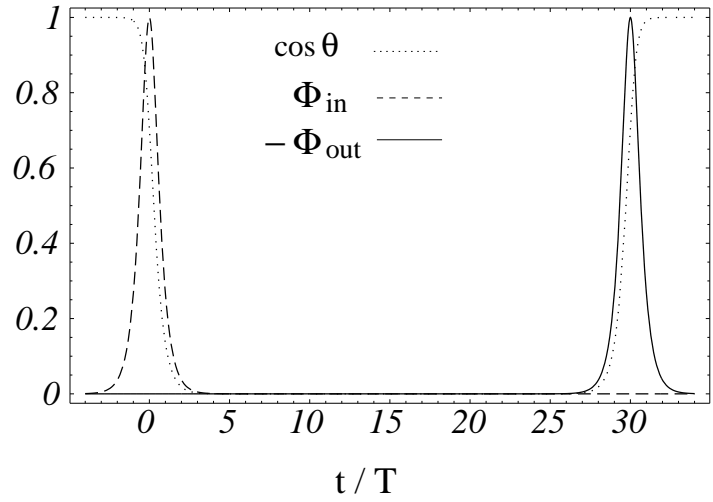,width=9.5cm}}
 \vspace*{2ex}
 \caption{Input and output wave functions for hyperbolic secant
input wave packet $\Phi_1$, $\gamma T=4$ and optimized
$\cos\theta(t)$. At $t\approx 30 T$ $\cos\theta(t)$ is time
reversed to release photon wave packet.  
$T$ characterizes the time unit. 	\label{fig5.fig}
}
\end{figure}


\begin{figure}[ht]
 \centerline{\epsfig{file=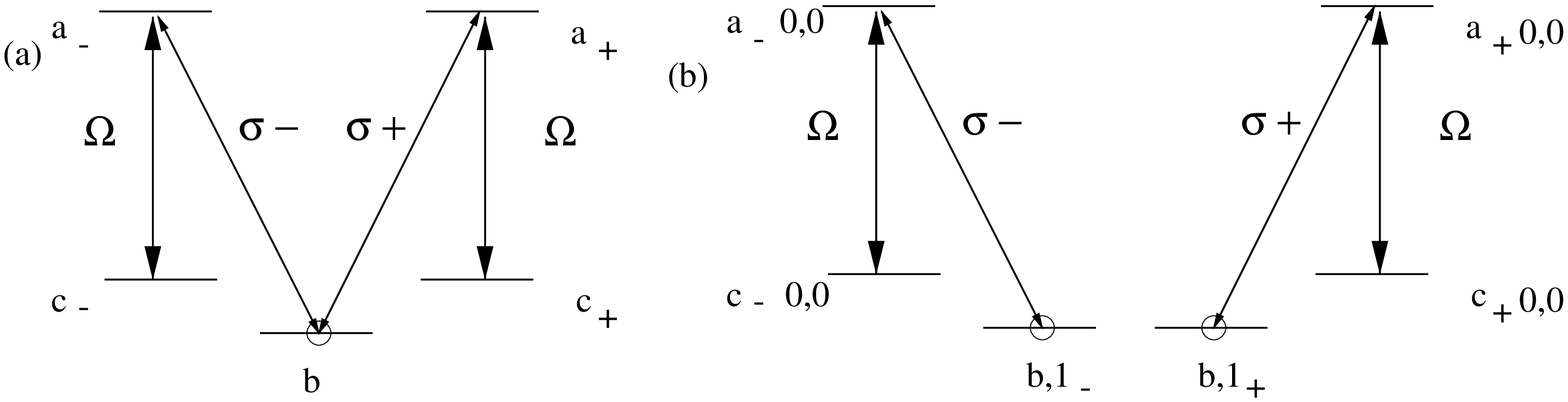,width=12 cm}}
 \vspace*{2ex}
 \caption{(a) Prototype of a multi-state atom for storing polarization 
states of quantum field. (b) Interaction of single-photon wave-packets of 
different polarizations with collective excitations. 
   	\label{fig6.fig}
}
\end{figure}



\begin{thebibliography}{99}


\bibitem{cel} M. O. Scully, Phys. Rev. Lett. {\bf 55}, 2802 (1985);
W. Schleich and M. O. Scully, Phys. Rev. A {\bf 37}, 
1261 (1988);  M. P. Winters, J. L. Hall and P. Toscheck, Phys. Rev.
A {\bf 65}, 3116 (1990).

\bibitem{scullybook}
 See for example chapter VII of
 M.\ O.\ Scully and M.\ S.\ Zubairy,
   {\em ``Quantum Optics''}, Cambrige University Press 1997.


\bibitem{eit} see e.g.: S.~E.~Harris, Physics Today {\bf 50}, 36 (1997)
and references therein. 

\bibitem{lwi} O. Kocharovskaya and Y. Khanin, JETP Lett. {\bf 48},
630 (1988); S. E. Harris, Phys. Rev. Lett. {\bf 62}, 1033 (1989);
M. O. Scully, S.-Y. Zhu, and A. Gavrielides, {\it ibid.} {\bf 62},
2813 (1989); A. S. Zibrov et al. {\it ibid.} {\bf 75}, 1499 (1995);
G. G. Padmabandu et al.  {\it ibid.} {\bf 76}, 2053 (1996).

\bibitem{iwl} 
D. A. Cardimona, M. G. Raymer, and C. R. Stroud, J. Phys. B
{\bf 15}, 55 (1982); A. Imamo\u glu, Phys. Rev. A {\bf 40}, 2835 (1989); 
S.-Y. Zhu and M. O. Scully, Phys. Rev. Lett. {\bf 76}, 388
(1996); H. R. Xia, C. Y. Ye, and S. Y. Zhu, Phys. Rev. Lett. {\bf 77}, 1032
(1996).

\bibitem{magnetometry} M. O. Scully and M. Fleischhauer, Phys. Rev. Lett.
{\bf 69}, 1360 (1992); S. Brandt, A. Nagel, R. Wynands and D. Meschede,
Phys. Rev. A {\bf 56}, R1063 (1997); 
 V.~A.~Sautenkov, M.~D.~Lukin, C.~J.~Bednar, G.~R.~Welch,
M.~Fleischhauer, V.~L.~Ve\-li\-chans\-ky and M.~O.~Scully,
(preprint: quant-ph/9904032).

\bibitem{lukin97prl}
 M.\ D.\ Lukin, M.\ Fleischhauer, A.\ S.\ Zibrov, H.\ G.\ Robinson,
	V.\ L.\ Velichansky, L.\ Hollberg, and M.\ O.\ Scully,
    \prl \textbf{79}, 2959 (1997).

\bibitem{index} M. O. Scully, Phys. Rev. Lett. {\bf 67}, 1855 (1991); 
A. S. Zibrov, M. D. Lukin, L. Hollberg, D. E. Nikonov, M. O. Scully,
H. G. Robinson, and V. L. Velichansky, Phys. Rev. Lett. {\bf 76}, 3935
(1996). 

\bibitem{nonlinear} S. E. Harris, J. E. Field and 
 A. Imamo\u glu, Phys. Rev. Lett.
{\bf 64}, 1107 (1990); K. Hakuta, L. Marmet and B. P. Stoicheff,
Phys. Rev. Lett. {\bf 66}, 596 (1991);
M. Jain, H. Xia, G. Yin, A. J. Merriam und
S. E. Harris, Phys. Rev. Lett. {\bf 77}, 4326 (1996);
 M.~D.~Lukin, P.~Hemmer, M.~L\"offler and M.~O.~Scully, 
Phys. Rev. Lett. {\bf 81}, 2675 (1998); M. Fleischhauer,
M. D. Lukin, A. B. Matsko, and M. O. Scully, Phys. Rev. Lett.
{\bf 82}, 1847 (1999). 


\bibitem{Hau99} 
 L.\ V.\ Hau, S.\ E.\ Harris, Z.\ Dutton, and C.\ H.\ Behroozi,
  Nature \textbf{397}, 594 (1999).

\bibitem{Kash99} M. M. Kash, V. A. Sautenkov, A. Zibrov, L. Hollberg, 
G. R. Welch, M. D. Lukin, Y. Rostovtsev, E. Fry, and M. O. Scully,
 Phys. Rev. Lett.
{\bf 82}, 5229 (1999). 

\bibitem{budker99} D. Budker, D. F. Kimball, S. M. Rochester, and V. V. 
Yashchuk,
 Phys. Rev. Lett.
{\bf 82}, (1999), in press.


\bibitem{Parkins93} A. S. Parkins, P. Marte, P. Zoller and H. J. Kimble,
Phys. Rev. Lett. {\bf 71}, 3095 (1993).

\bibitem{STIRAP}  J. Oreg, F. T. Hioe, and J. H. Eberly, Phys. Rev. {\bf A} 
29, 690 (1984);  U. Gaubatz, P. Rudecki, M. Becker, S. Schiemann, M. 
K\"{u}lz, and K. Bergmann, Chem. Phys. Lett. {\bf 149}, 463 (1988);
 K. Bergmann, H. Theuer, and B. W. Shore,  Rev. Mod. Phys. 
{\bf 70}, 1003 (1998).  


\bibitem{Cirac97} J. I. Cirac, P. Zoller, H. J. Kimble, and H. Mabuchi,
 Phys. Rev. Lett. {\bf 78}, 3221 (1997).

\bibitem{Enk97} S. J. van Enk, J. I. Cirac, and P. Zoller,
 Phys. Rev. Lett. {\bf 78}, 4293 (1997).

\bibitem{Pellizzari97} T. Pellizzari, Phys. Rev. Lett. {\bf 79}, 5242 (1997).

\bibitem{Enk98} S. J. van Enk, H. J. Kimble, J. I. Cirac, and P. Zoller,
Phys. Rev. A {\bf 59}, 2659 (1999).

\bibitem{Law97} C. K. Law and H. J. Kimble, J. Mod. Opt. {\bf 44}, 2067 (1997).

\bibitem{Imamoglu94} A. Imamo\u glu and Y. Yamamoto, Phys. Rev. Lett. {\bf 72},
210 (1994); F. De Martini et al., {\it ibid.} {\bf 76}, 900 (1996).

\bibitem{Gheri98} K. M. Gheri, C. Saavedra, P. T\"orm\"a, J. I. Cirac, and
P. Zoller,  Phys. Rev. A {\bf 58}, R2627 (1998).



\bibitem{Pellizzari95} T. Pellizzari, S. A. Gardiner, J. I. Cirac und
P. Zoller, Phys. Rev. Lett. {\bf 75}, 3788 (1995).

\bibitem{kimble} H. J. Kimble, Physica Scripta, {\bf 76}, 127 (1998). 

\bibitem{OL98} M. D. Lukin, M. Fleischhauer, M. O. Scully, and
V. L. Velichansky, Opt. Lett. {\bf 23}, 295 (1998).

\bibitem{polzhik1} A.~Kuzmich, K.~M\o lmer, and E.~S.~Polzik, 
Phys. Rev. Lett. {\bf 79}, 4782 (1997).

\bibitem{polzhik2} J.~Hald, J.~L.~S\o rensen, C.~Schori, and E.~S. Polzik, 
Phys. Rev. Lett. {\bf 83}, 1319 (1999).

\bibitem{werner} M. Werner, and A. Imamo\u glu, preprint quant-ph/9902005.


\bibitem{meystre} E. S. Lee, C. Geckeler, J. Heurich, A. Gupta, Kit-Iu Cheong, S. Secrest and P. Meystre, "Dark states of dressed Bose-Einstein condensates'', Phys. Rev. A, to be published. 

\bibitem{Cohen-Tannoudji} C. Cohen-Tannoudji and S. Reynaud, 
J. Phys. B {\bf 10}, 2311 (1977).

\bibitem{Siegmann} A. Siegmann, {\it Lasers}, (University Science Books,
Mill Valley, CA, 1986)



\end{thebibliography}
\end{document}